\documentclass[reprint,aps,graphicx,pre,showpacs]{revtex4-1}
\usepackage{amsmath,amssymb,amsfonts}   
\usepackage{graphicx,bm}
\usepackage{paralist}
\usepackage{epstopdf}
\usepackage{graphics} 
\usepackage[colorlinks=true]{hyperref}
\hypersetup{urlcolor=blue, citecolor=red}
\usepackage[latin1]{inputenc}

\renewcommand{\theequation}{\arabic{section}.\arabic{equation}}

\newcommand{\R}{{\mathbb R}}

\newcommand{\calU}{{\mathcal U}}

\newcommand{\neb}{\bar{\branch}} 

\newcommand{\targT}{\mathcal{X}} 
\newcommand{\branch}{\mathcal{I}}
\newcommand{\flx}{\mathcal{J}}

\newcommand{\uoo}{\psi}

\newcommand{\e}{\mathrm{e}}

\newcommand{\tu}{\widetilde{p}}

\newcommand{\x}{\mathbf{x}}

\renewcommand{\P}{\mathbb{P}}

\newcommand{\n}{\mathbf n}

\newcommand{\pcb}[1]{\textcolor{blue}{#1}}

\begin{document}

\title{First-passage processes and the target-based accumulation of resources}

\author{Paul C. Bressloff}
\address{Department of Mathematics, University of Utah 155 South 1400 East, Salt Lake City, UT 84112}

\begin{abstract}

Random search for one or more targets in a bounded domain occurs widely in nature, with examples ranging from animal foraging to the transport of vesicles within cells. Most theoretical studies take a searcher-centric viewpoint, focusing on the first passage time (FTP) problem to find a target. This single search-and-capture event then triggers a downstream process or provides the searcher with some resource such as food. In this paper we take a target-centric viewpoint, by considering the accumulation of resources in one or more targets due to multiple rounds of search-and-capture events combined with resource degradation; whenever a searcher finds a target it delivers a resource packet to the target, after which it escapes and returns to its initial position. The searcher is then resupplied with cargo and a new search process is initiated after a random delay. It has previously been show how queuing theory can be used to derive general expressions for the steady-state mean and variance of the resulting resource distributions. Here we apply the theory to some classical FPT problems involving diffusion in simple geometries with absorbing boundaries, including concentric spheres, wedge domains, and branching networks. In each case, we determine how the resulting Fano factor depends on the degradation rate, the delay distribution, and various geometric parameters. We thus establish that the Fano factor can deviate significantly from Poisson statistics and exhibits a non-trivial dependence on model parameters, including non-monotonicity and crossover behavior. This indicates the non-trivial nature of the higher-order statistics of resource accumulation.
\end{abstract}

\maketitle

\section{Introduction}

Random search strategies are found throughout the natural world as a way of efficiently searching for one or more targets of unknown location. Examples include animals foraging for food or shelter 
\cite{Bell91,Bartumeus09,Viswanathan11}, proteins searching for particular sites on DNA \cite{Berg81,Halford04,Coppey04,Lange15}, biochemical reaction kinetics \cite{Loverdo08,Benichou10}, motor-driven intracellular transport of vesicles \cite{Bressloff13,Maeder14,Bressloff15}, and cytoneme-based morphogen transport \cite{Kornberg14,Stanganello16,Zhang19,Bressloff19}. Most theoretical studies of these search processes take a searcher-centric viewpoint, focusing on the first passage time (FTP) problem to find a target. This single search-and-capture event then triggers a downstream process or provides the searcher with some resource such as food or shelter. An alternative approach is to take a target-centric viewpoint, whereby one keeps track of the accumulation of resources in the targets due to multiple rounds of search-and-capture events. In this case, whenever a searcher finds a target it delivers a resource packet to the target, after which it escapes and returns to its initial position. The searcher is then resupplied with cargo and a new search process is initiated. 

We have recently shown \cite{Bressloff19,Bressloff20a,Bressloff20q,Bressloff20A} how the steady-state distribution of resources accumulated by one or more targets can be determined by reformulating a search-and-capture model as a queuing process \cite{Takacs62,Liu90}. Renewal theory can then be used to calculate moments of the distribution of resources in steady state, which are expressed in terms of the Laplace transform of the probability fluxes into the set of targets under a single round of search-and-capture. The probability fluxes correspond to the (conditional) FPT densities of the search process. Our previous work has mainly focused on one-dimensional search problems and the effects of stochastic resetting. However, the basic framework applies to any search process. Therefore, in this paper, we apply the theory to a variety of classical FPT problems, involving diffusion in simple geometries with one or more absorbing boundaries, including concentric spheres, wedge domains, and branching networks. In each case, we first solve the FPT problem or state results from the literature. Identifying each absorbing boundary with a corresponding target, we then use the solution to the FPT problem to determine the steady-state mean and variance of the distribution of resources. Finally, we investigate the dependence of the corresponding Fano factor (variance/mean) on the degradation rate, distribution of delays, and various geometric parameters. In particular, we establish that significant deviations from Poisson statistics can occur, with the Fano factor exhibiting non-monotonic and crossover behavior. The general theoretical framework is introduced in section II and then applied to a pair of concentric spheres (section III), a wedge domain (section IV), a network with a single side-branch (section V), and a semi-infinite Cayley tree (section VI).

\setcounter{equation}{0}
\section{Accumulation of resources under multiple search-and-capture events}

\begin{figure}[t!]
\centering
\includegraphics[width=5cm]{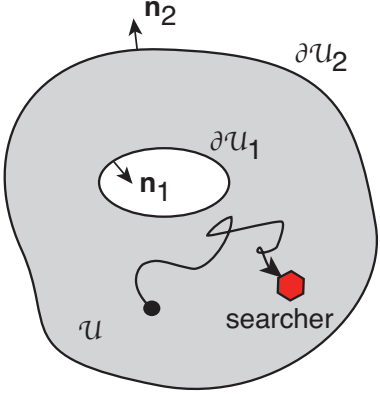} 
\caption{Brownian particle searching in a bounded domain ${\mathcal U}$ with absorbing interior and exterior boundaries $\partial\calU_k$, $k=1,2$. Here $\n_{1,2}$ denote normals to the boundaries.}
\label{fig1}
\end{figure}

Consider a bounded domain ${\mathcal U}\subset \R^d$ with an interior boundary $\partial\calU_1$ and an exterior boundary $\partial\calU_2$, see Fig. \ref{fig1}. Suppose that both boundaries are totally absorbing and thus act as a pair of targets. (One could also consider the simpler case of a single target by taking one of the boundaries to be reflecting, or include multiple interior targets.) Suppose that a particle (searcher) is subject to Brownian motion in $\calU$. The probability density $p(\x,t|\x_0)$ for the particle to be at position $\x$ at time $t$, having started at $\x_0$, evolves according to the diffusion equation 
\begin{equation}
 \label{1J}
\frac{\partial p(\x,t|\x_0)}{\partial t}=D\nabla^2p(\x,t|\x_0)=-\nabla\cdot {\mathbf J}(\x,t|\x_0),
\end{equation}
where ${\mathbf J}=-D\nabla p$ is the probability flux. This is supplemented by the boundary condition
\begin{equation}
 p(\x,t|\x_0)=0,\ \x \in  \bigcup_{j=1}^2\partial{\mathcal U}_j,
\end{equation}
and the initial condition $p(\x,0|\x_0)=\delta(\x-\x_0)$. 

Let ${\mathcal T}_k(\x_0)$ denote the FPT that the particle is captured by the $k$-th target, with ${\mathcal T}_k(\x_0)=\infty$ indicating that it is not captured.
Define $\Pi_k(\x_0,t)$ to be the probability that the particle is captured by the $k$-th target after time $t$, given that it started at $\x_0$:
\begin{equation}
\label{Pi}
\Pi_k(\x_0,t)=\P[t<{\mathcal T}_k(\x_0)<\infty ]=\int_t^{\infty} J_k(\x_0,t')dt',
\end{equation}
where
\begin{equation}
\label{Jk}
J_k(\x_0,t)=\int_{\partial \calU_k} {\mathbf J} (\sigma,t|\x_0)\cdot {\bf n} d\sigma.
\end{equation}
Note that the normal $\n$ to the boundary $\partial U_k$ is always taken to point from inside to outside the search domain, see Fig. \ref{fig1}. Moreover, differentiating Eq. (\ref{Pi}) and taking Laplace transforms implies that
\begin{equation}
\label{PiLT}
s\widetilde{\Pi}_k(\x_0,s)-\pi_k(\x_0)=-\widetilde{J}_k(\x_0,s).
\end{equation}
 The splitting probability $\pi_k(\x_0)$ and conditional MFPT $T_k(\x_0)$ for the particle to be captured by the $k$-th target are then
\begin{equation}
\label{pi}
\pi_k(\x_0)  =\Pi_k(\x_0,0)= \int_0^\infty J_k(\x_0,t) dt=\widetilde{J}_k(\x_0,0),
\end{equation}
and
\begin{align}
T_k(\x_0) &= \mathbb{E}[{\mathcal T}_k | {\mathcal T}_k < \infty]=\frac{1}{\pi_k(\x_0)}\int_0^\infty \Pi_k(\x_0,t) dt \nonumber \\
&=-\pcb{\frac{1}{\pi_k(\x_0)}}\left . \frac{\partial \widetilde{J}_k(\x_0,s)}{\partial s}\right |_{s=0}.
\end{align}
We will assume that $\sum_{k=1,2}\pi_k(\x_0)=1$, which implies that the particle is eventually captured by a target with probability one. Finally, note that integrating equation (\ref{1J}) with respect to $\x$ and $t$ implies that the survival probability up to time $t$ is
\begin{align}
\label{q00}
Q(\x_0,t)&= \int_{\calU}p(\x,t|\x_0)d\x=\sum_{k=1,2}\Pi_k(\x_0,t).
\end{align}
In Laplace space,
\begin{align}
\label{qLT}
s\widetilde{Q}(\x_0,s)&=1-\sum_{k=1,2}\widetilde{J}_k(\x_0,s).
\end{align}

\begin{figure}[t!]
\begin{center} 
\includegraphics[width=8.5cm]{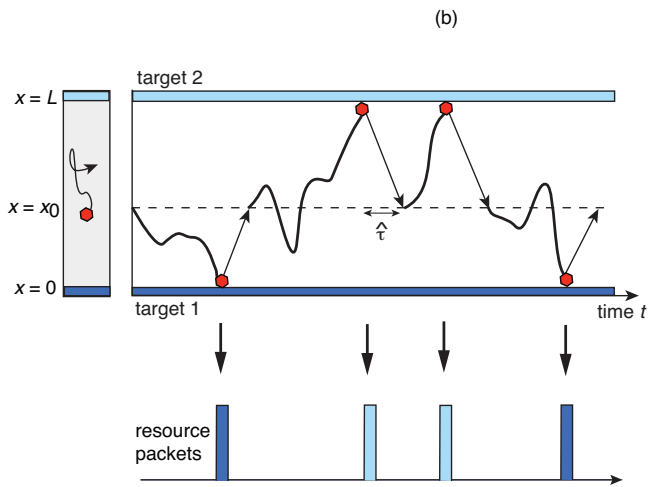} 
\caption{Multiple search-and-capture events for a particle searching for a pair of exterior targets located at the ends of the finite interval $x\in [0,L]$. Each time the particle reaches (finds) a target it delivers a discrete packet of resources (capture event) and then returns to its initial position $x_0$, $0<x_0<L$, where it is loaded with another packet and the process repeats. The delay time $\widehat{\tau}$ between a capture event and initiation of a new search is generated from a waiting time density $\rho(\widehat{\tau})$. (The lines with arrows do not represent actual trajectories.) The sequence of capture events results in an accumulation of resource packets within each target, which is counteracted by degradation at some rate $\gamma$.}
\label{fig2}
\end{center}
\end{figure}

Now suppose that, rather than being permanently absorbed or captured by a target on the boundary, the particle delivers a discrete packet of some resource to the target and then returns to $\x_0$, initiating another round of search-and-capture. We will refer to the delivery of a single packet as a capture event. The sequence of events resulting from multiple rounds of search-and-capture leads to an accumulation of packets within the targets, which we assume is counteracted by degradation at some rate $\gamma$. This is illustrated in Fig. \ref{fig2} for the finite interval.
We will assume that the total time for the particle to unload its cargo, return to $\x_0$ and start a new search process is given by the random variable $\widehat{\tau}$, which for simplicity is taken to be independent of the particular absorbing boundary. (This is reasonable if the sum of the mean loading and unloading times is much larger than a typical return time.) If $\rho(\widehat{\tau})$ denotes the waiting time density of the delay $\widehat{\tau}$, then the conditional first passage time density for delivery of a packet to the $j$-th target is
\begin{align}
 {\mathcal F}_{j}(\Delta) &=\int_0^{\Delta} f_{j}(t) \rho(\Delta-t)dt,
\end{align}
where $f_{j}(t)=J_j(t)/\pi_j$ is the conditional first passage time density for a single search-and-capture event without delays. (For notational simplicity, we drop the explicit dependence on the initial position $\x_0$.) Laplace transforming the convolution equation then yields
\begin{equation}
\label{calF}
\widetilde{{\mathcal F}}_{j}(s)=\widetilde{f}_{j}(s)\widetilde{\rho}(s).
\end{equation}

As we have shown elsewhere \cite{Bressloff19,Bressloff20a,Bressloff20q,Bressloff20A}, the steady-state distribution of resources accumulated by the targets can be determined by reformulating the model as a queuing process \cite{Takacs62,Liu90}. Here we simply state the results for the steady-state mean and variance. Let $M_k$ be the steady-state number of resource packets in the $k$-th target. The mean is then
\begin{equation} \label{mean}
\overline{M}_k  =  \frac{\pi_k}{\gamma \sum_{j=1,2} \pi_{j} (T_j +\tau_{\rm cap} )}=\frac{\pi_k}{\gamma(T+\tau_{\rm cap})},
\end{equation}
where $\tau_{\rm cap}=\int_0^{\infty}\rho(\tau)d\tau$ is the mean loading/unloading time and $T=\sum_{j=1,2}\pi_jT_j$ is the unconditional MFPT.
Eq. (\ref{mean}) is consistent with the observation that $T +\tau_{\rm cap} $ is the mean time for one successful delivery 
of a packet to any one of the targets and initiation of a new round of search-and-capture. Hence, its inverse is the mean rate of capture events and $\pi_k$ is the fraction that are delivered to the $k$-th target (over many trials). (Note that Eq. (\ref{mean}) is known as Little's law in the queuing theory literature \cite{Little61} and applies more generally.) The dependence of the mean $\overline{M}_k$ on the target label $k$ specifies the steady-state allocation of resources across the set of targets. It will depend on the details of the particular search process (\ref{1J}), the geometry of the domain $\calU$, the initial position $\x_0$, and the rate of degradation $\gamma$. Similarly, 
the variance of the number of resource packets is
\begin{align}
\label{var0}
 \mbox{Var}[M_k]=\overline{M}_k\left [ \frac{\pi_{k}\widetilde{\mathcal F}_{k}(\gamma)}{1-\sum_{j=1,2}\pi_{j} \widetilde{\mathcal F}_{j}(\gamma)}+1-\overline{M}_k
\right ]. 
\end{align}
Finally, noting that $\pi_{k}\widetilde{\mathcal F}_{k}(\gamma)=\widetilde{\rho}(\gamma)\widetilde{J}_{k}(\gamma)$ and using Eq. (\ref{qLT}) yields
\begin{align}
\label{var}
 \mbox{Var}[M_k]=\overline{M}_k\left [ \frac{\widetilde{\rho}(\gamma)\widetilde{J}_{k}(\gamma) }{1-\widetilde{\rho}(\gamma)[1-\gamma \widetilde{Q}(\gamma)]}+1-\overline{M}_k
\right ]. 
\end{align}
The above results straightforwardly extend to $N$ targets, $N\geq 1$, with $\pi_1=1$ if $N=1$.

Both the mean and variance vanish in the fast degradation limit $\gamma \rightarrow \infty$, since resources delivered to the targets are immediately degraded so that there is no accumulation. On the other hand, in the limit of slow degradation ($\gamma \rightarrow 0$), the mean and variance both become infinite. (There is no stationary state when $\gamma=0$.) Rather than working with the variance, however, it is more convenient to consider the Fano factor
\begin{equation}
\label{FF}
 FF_{k}=\frac{\mbox{Var}[M_{k}]}{\overline{M}_{k}}= 1+ \frac{\widetilde{\rho}(\gamma)\widetilde{J}_{k}(\gamma) }{1-\widetilde{\rho}(\gamma)[1-\gamma \widetilde{Q}(\gamma)]}-\overline{M}_k. 
 \end{equation}
It immediately follows that
\begin{equation}
\label{cong}
 \lim_{\gamma \rightarrow \infty} FF_{k}=1.
\end{equation}
The Fano factor is a natural quantity to consider in the case of a queuing process, since the latter is an example of a counting process that tracks discrete events. The best known example of a counting process is the Poisson process, which is Markovian and has a Fano factor of one. In applications this is often used as a baseline to characterize the level of noise in a counting process. For example, neural variability in experiments is typically specified in terms of the statistics of spike counts over some fixed time interval, and compared to an underlying inhomogeneous Poisson process \cite{Faisal08}. Often Fano factors greater than one are observed, indicative of some form of spike bursting \cite{Softky92}. Fano factors greater than unity are also a signature of protein bursting in gene networks \cite{Bose04,Collins05,Friedman06}.

One important point to emphasize is that we consider ensemble or trial averages in this paper. However, assuming that the stochastic process is ergodic, one would obtain identical statistics by observing temporal variations in target resources.
Finally, note that an alternative measure of noise is the coefficient of variation, which for the $k$-th target is defined according to
\begin{equation}
\label{CV}
 CV_{k}=\frac{\mbox{Var}[M_{k}]}{\overline{M}_{k}^2}. 
 \end{equation}
Given the asymptotic behavior of $FF_k$, we find that
\[ \lim_{\gamma \rightarrow 0} CV_{k}=0,\quad \lim_{\gamma \rightarrow \infty} CV_{k}=\infty.\]

\setcounter{equation}{0}
\section{Diffusive search between concentric spheres}

As our first example, consider the classical problem of diffusive search between two concentric $d$-dimensional spheres of radii $R_1$ and $R_2$, respectively, with $R_2>R_1$, see Fig. \ref{fig4} and Ref. \cite{Redner}. In the 1D case this reduces to the problem of diffusive search in a finite interval of length $L=R_2-R_1$ with absorbing boundaries at the ends $x=0,L$. 

 \begin{figure}[t!]
\begin{center} 
\includegraphics[width=7cm]{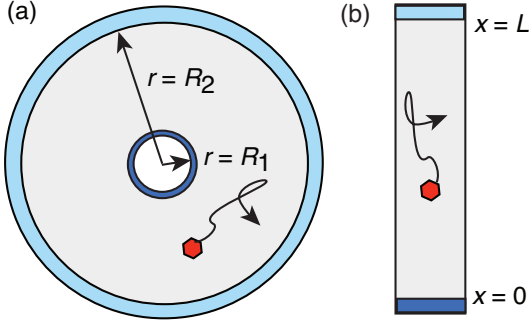} 
\caption{(a) Search domain consisting of the region between a pair of concentric $d$-dimensional spheres of radii $R_1,R_2$. (b) In 1D the search domain reduces to a finite interval of length $L=R_2-R_1$ with absorbing boundaries at $x=0,L$.}
\label{fig3}
\end{center}
\end{figure}

\subsection{Splitting probabilities and unconditional MFPT}

Using radial symmetry, Eq. (\ref{1J}) reduces to the 1D equation
\begin{subequations}
\label{2DJ}
\begin{equation}
\frac{\partial p(r,t|r_0)}{\partial t}=D\left [\frac{\partial^2 p(r,t|r_0)}{\partial r^2}+\frac{d-1}{r}\frac{\partial p(r,t|r_0)}{\partial r}\right ],
\end{equation}
where $r$ represents the radial coordinate. This is,
supplemented by the boundary conditions
\begin{align}
 p(R_1,t|r_0)=0=p(R_2,t|r_0),
\end{align}
\end{subequations}
and the initial condition 
\begin{equation}
\label{init}
p(r,0|r_0)=\frac{\delta(r-r_0)}{\Omega_dr_0^{d-1}},
\end{equation}
with $\Omega_d$ the surface area of the $d$-dimensional unit sphere. That is, the initial condition is uniformly distributed around the spherical surface of radius $r_0$. 
Laplace transforming the radial diffusion equation gives
\begin{subequations}
\label{2LT}
\begin{align}
&D\left [\frac{\partial^2 \widetilde{p}(r,s|r_0)}{\partial r^2}+\frac{d-1}{r}\frac{\partial\widetilde{p}(r,s|r_0)}{\partial r}\right ]-s \widetilde{p}(r,s|r_0)\nonumber \\
&\qquad =-\frac{\delta(r-r_0)}{\Omega_dr_0^{d-1}},
\end{align}
supplemented by the boundary conditions
\begin{align}
 \widetilde{p}(R_1,s|r_0)=0=\widetilde{p}(R_2,s|r_0).
\end{align}
\end{subequations}
The solution of Eq. (\ref{2LT}) is carried out in Ref. \cite{Redner}, and one finds
\begin{align}
\widetilde{p}(r,s|r_0)&=\frac{(rr_0)^{\nu}}{D\Omega_d}\frac{C_{\nu}(r,R_1;s)C_{\nu}(r_0,R_2;s)}{C_{\nu}(R_1,R_2;s)},\ r<r_0,\nonumber\\
\widetilde{p}(r,s|r_0)&=\frac{(rr_0)^{\nu}}{D\Omega_d}\frac{C_{\nu}(r_0,R_1;s)C_{\nu}(r,R_2;s)}{C_{\nu}(R_1,R_2;s)},\ r>r_0,
\end{align}
where $\nu=1-d/2$,
\begin{align*}
& C_{\nu}(a,b;s)\\
&=I_{\nu}(\sqrt{s/D}a)K_{\nu}(\sqrt{s/D}b)-I_{\nu}(\sqrt{s/D}b)K_{\nu}(\sqrt{s/D}a),
\end{align*}
and $I_{\nu}$ and $K_{\nu}$ are the modified Bessel functions of the first and second kind, respectively. 

\begin{figure*}[t!]
\begin{center} 
\includegraphics[width=15cm]{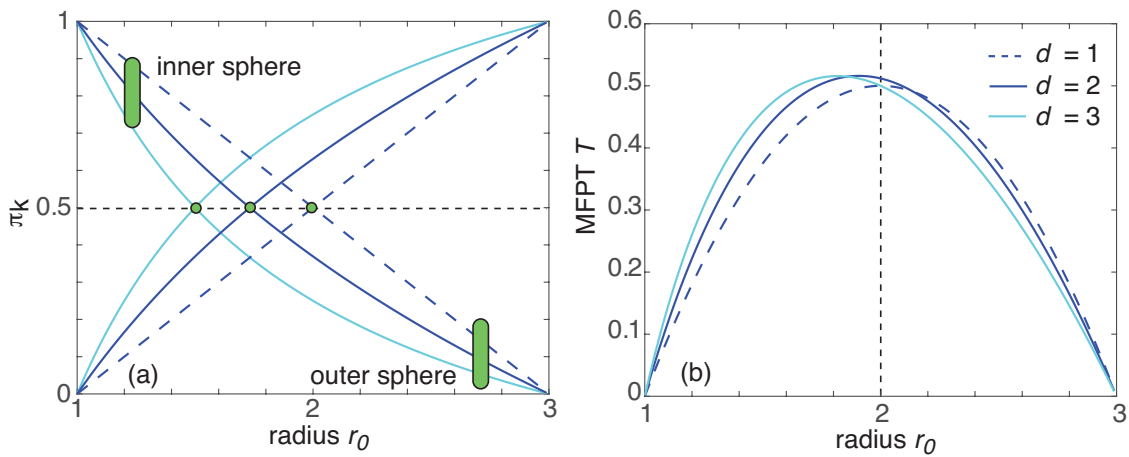} 
\caption{Plots of (a) splitting probabilities $\pi_{k}$ of inner ($k=1$) and outer ($k=2$) spheres and (b) unconditional MFPT $T$ as a function of the initial position $r_0$ for $d=1,2,3$. Other parameters are $D=1,R_1=1,R_3=3$. Green dots in (a) indicate the initial location where the splitting probabilities in the two targets are balanced.}
\label{fig4}
\end{center}
\end{figure*}

The FPT properties can now be obtained by integrating the probability flux over the surface boundary of each sphere and taking Laplace transforms. Hence, for the inner target of radius $R_1$, 
\begin{subequations}
\label{tilJ}
\begin{align}
\widetilde{J}_1(r_0,s)&=D\Omega_d R_1^{d-1}\left . \frac{\partial\widetilde{p}(r,s|r_0)}{\partial r}\right |_{r=R_1}\nonumber \\
&=\left (\frac{r_0}{R_1}\right )^{\nu}\frac{C_{\nu}(r_0,R_2;s)}{C_{\nu}(R_1,R_2;s)}.
\end{align}
Similarly, for the outer target of radius $R_2$,
\begin{align}
\widetilde{J}_2(r_0,s)&=-D\Omega_d R_2^{d-1}\left . \frac{\partial\widetilde{p}(r,s|r_0)}{\partial r}\right |_{r=R_2}\nonumber \\
&=-\left (\frac{r_0}{R_2}\right )^{\nu}\frac{C_{\nu}(r_0,R_1;s)}{C_{\nu}(R_1,R_2;s)}.
\end{align}
\end{subequations}
We have used the Bessel function identities 
\begin{align*}
I_{\nu}'(r)&=-\frac{\nu}{r}I_{\nu}(r)+I_{\nu-1}(r),\\ K_{\nu}'(r)&=-\frac{\nu}{r}K_{\nu}(r)-K_{\nu-1}(r),
\end{align*}
which imply that
\[ \left .\frac{\partial C_{\nu}(r,y;s)}{\partial r}\right |_{y=r}=I_{\nu-1}(r)K_{\nu}(r)+I_{\nu}(r)K_{\nu-1}(r)=\frac{1}{r}.\]

It follows from Eq. (\ref{pi}) that the splitting probability $\pi_k(r_0)$, $k=1,2$, for being captured by the inner and outer spherical boundaries is given by
\begin{equation}
\pi_k(r_0)=\lim_{s\rightarrow 0} \widetilde{J}_k(r_0,s).
\end{equation}
Hence \cite{Redner},
\begin{align}
\label{pi2}
\pi_2(r_0)=\left \{ \begin{array}{cc} \frac{\displaystyle 1-(R_1/r_0)^{d-2}}{\displaystyle 1-(R_1/R_2)^{d-2}} & d\neq 2\\ & ,\\
\frac{\displaystyle \ln(R_2/r_0)}{\displaystyle \ln(R_2/R_1)}& d=2 \end{array} \right .
\end{align}
with $\pi_1(r_0)=1-\pi_2(r_0)$. The Laplace transform $\widetilde{\Pi}_k(r_0,r)$ is then obtained from Eq. (\ref{PiLT}). Finally, the Laplace transform of the survival probability is
\begin{align}
&\widetilde{Q}(r_0,s)=\widetilde{\Pi}_1(r_0,s)+\widetilde{\Pi}_2(r_0,s)\\
&=\frac{1}{s}\left [1-\widetilde{J}_1(r_0,s)-\widetilde{J}_2(r_0,s)\right ]\nonumber \\
&=\frac{1}{s}\left [1+\left (\frac{r_0}{R_2}\right )^{\nu}\frac{C_{\nu}(r_0,R_1)}{C_{\nu}(R_1,R_2)}-\left (\frac{r_0}{R_1}\right )^{\nu}\frac{C_{\nu}(r_0,R_2)}{C_{\nu}(R_1,R_2)}\right ],\nonumber
\end{align}
and the conditional FPTs are
\begin{equation}
\label{eq7:LTf}
\widetilde{f}_k(r_0,s)=\frac{\widetilde{J}_k(r_0,s)}{\pi_k(r_0)}.
\end{equation}

In Fig. \ref{fig4} we plot the splitting probabilities $\pi_{k}$, $k=1,2$, and the unconditional MFPT $T$ as a function of the initial position $r_0$ for $d=1,2,3$. As expected, $\pi_{2}(R_1)=0$, $\pi_{2}(R_2)=1$ and  $T(R_{1,2})=0$. The 1D case has a reflection symmetry about the midpoint $(R_1+R_2)/2$, that is, $\pi_{k}(r_0)=\pi_{k}(L-r_0)$ for $L=R_2-R_1$ and similarly for $T$. On the other hand, in higher dimensions the curves are skewed towards the inner sphere, since it has a smaller surface area and is thus a less effective trap compared to the outer sphere.

\subsection{Mean and Fano factor of resource distribution}

\begin{figure*}[t!]
\begin{center} 
\includegraphics[width=15cm]{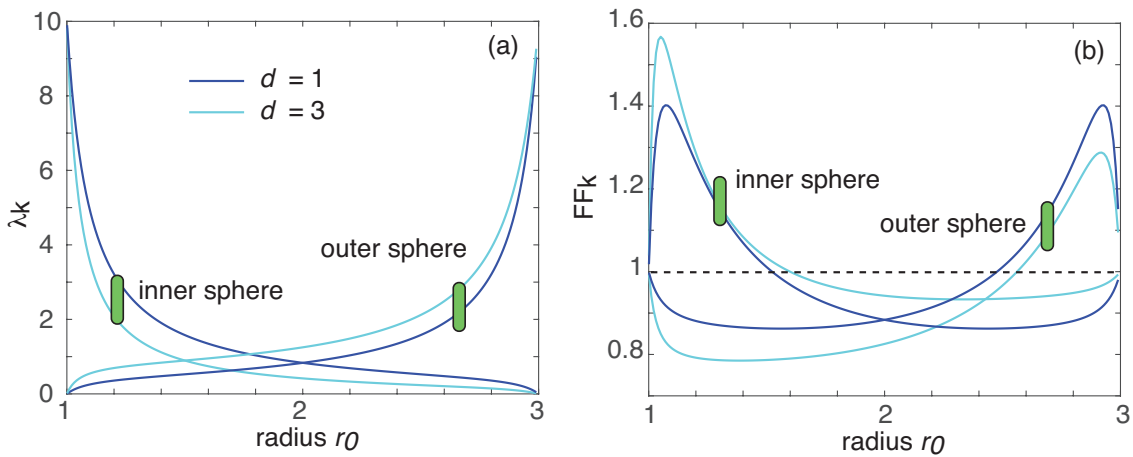} 
\caption{Plot of steady-state (a) accumulation rate $\lambda_{k}=\gamma \overline{M}_k$ and (b) Fano factor $FF_{k}$ in the $k$-th sphere as a function of the initial position $r_0$ for dimensions $d=1,3$, $\gamma=1$, $\tau_{\rm cap}=0.1$ and $\beta=1$. Other parameters are as in Fig. \ref{fig4}.}
\label{fig5}
\end{center}
\end{figure*}

\begin{figure*}[t!]
\begin{center} 
\includegraphics[width=15cm]{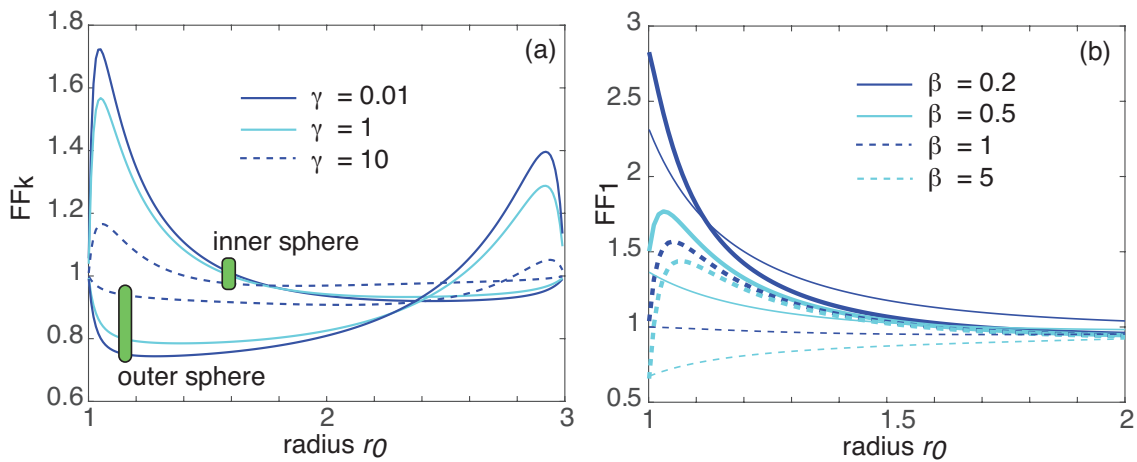} 
\caption{(a) Plot of steady-state Fano factor $FF_{k}$ in the $k$-th sphere as a function of the initial position $r_0$ for various degradation rates $\gamma$ and $\tau_{\rm cap}=0.1,\beta=1,d=3$. (b) Plot of steady-state Fano factor $FF_{1}$ in the inner sphere as a function of the initial position $r_0$ for various gamma distributions $\rho(\tau)$ and $\gamma =1$. Thick (thin) curves correspond to $\tau_{\rm cap}=0.1$ ($\tau_{\rm cap}=1$). Other parameters are as in Fig. \ref{fig5}.}
\label{fig6}
\end{center}
\end{figure*}

\begin{figure}[t!]
\begin{center} 
\includegraphics[width=8cm]{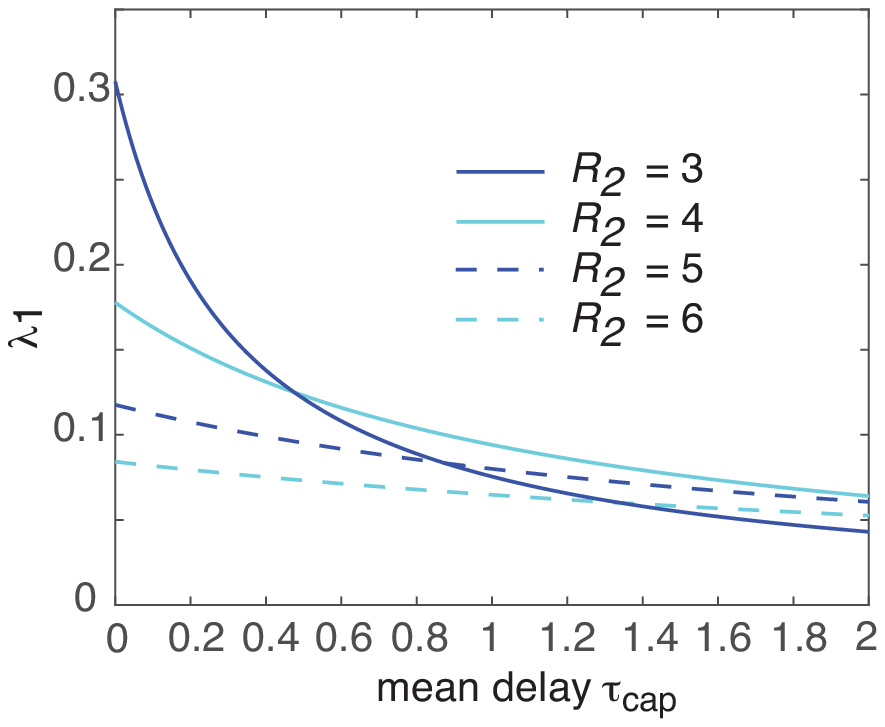} 
\caption{Plot of accumulation rate $\lambda_{1}$ in the inner sphere as a function of the mean delay $\tau_{\rm cap}$ for various outer radii, $d=3$, and $r_0=2.5$. Other parameters are as in Fig. \ref{fig5}.}
\label{fig7}
\end{center}
\end{figure}

We now use Eqs. (\ref{tilJ}), (\ref{pi2}) and (\ref{eq7:LTf}) to determine the steady-state mean and Fano factor of the resource distribution between the two targets, which are given by Eqs. (\ref{mean}) and (\ref{FF}), respectively. Since $T$ vanishes at the boundaries, it follows from Eq. (\ref{mean}) that the means $\overline{M}_{k}$ are singular at the boundaries in the absence of a capture delay. This reflects the fact that without any delays, the frequency at which the particle loads and unloads resources becomes unbounded, which is physically unrealistic. In order to remove this singular behavior, we take $\tau_{\rm cap}>0$. Since the Fano factor depends on $\widetilde{\rho}(s)$, we need to specify the density $\rho(\tau)$ for capture delays. For the sake of illustration, we consider the Gamma distribution
\begin{equation}
\label{gam}
\rho(\tau)=\frac{1}{\Gamma(\beta)a^{\beta}}\tau^{\beta-1}\e^{-\tau/a},
\end{equation}
with mean $\tau_{\rm cap}=\beta a$ and variance $\mbox{Var}[\tau]=\beta a^2$. This choice of $\rho(\tau)$ allows us to vary the mean and variance of $\rho(\tau)$ independently by varying $a$ and $\beta$. (The case $\beta=1$ corresponds to an exponential density). The corresponding Laplace transform is
\begin{equation}
\widetilde{\rho}(s)=\frac{1}{(1+s a)^{\beta}}=\frac{1}{(1+s \tau_{\rm cap}/\beta)^{\beta}}.
\end{equation}
We also assume that after each capture event, the particle returns to a random location on the circle of radius $r_0$, consistent with Eq. (\ref{init}). For the given FPT problem, we explore how $\overline{M}_k$ and $FF_k$ depend on model parameters such as the initial radius $r_0$, the dimension $d$, the degradation rate $\gamma$, and the delay parameters $\tau_{\rm cap},\beta$.

In Fig. \ref{fig5} we show sample plots of the steady-state accumulation rate $\lambda_{k}=\gamma \overline{M}_{k}$ and the Fano factor $FF_{k}$ as a function of the radius $r_0$ for $d=1,3$. The parameters of the Gamma distribution are $\tau_{\rm cap}=0.1$ and $\beta =1$. We find that the accumulation rate increases (decreases) with the dimension $d$ for the outer (inner) sphere and is a monotonic function of $r_0$. On the other hand, the Fano factor is a non-monotonic function of the initial location and its dependence on $d$ is the opposite of the accumulation rate. Moreover, the Fano factor tends to be greater than unity for initial locations close to a target, and less than unity at more distal locations, at least in the given parameter regime. In Fig. \ref{fig6} we further explore the dependence of the Fano factor on various model parameters. Fig. \ref{fig6}(a) shows how varying the degradation rate $\gamma$ changes the Fano factor. Consistent with the analysis of section II, $FF_k$ approaches unity for large $\gamma$, that is, the curves flatten. Moreover, the curves converge in the limit $\gamma \rightarrow 0$. Another result is that there is a crossover in the dependence of $FF_k$ on $\gamma$, whereby $FF_k$ is an increasing function of $\gamma$ for distal starting radii $r_0$ and a decreasing function of $\gamma$ for proximal $r_0$. Fig. \ref{fig6}(b) illustrates how the Fano factor depends on the parameters $\tau_{\rm cap}$ and $\beta$ of the gamma distribution (\ref{gam}). In general, we find that reducing $\beta$ for fixed $\tau_{\rm cap}$ (increasing the variance of the capture delay $\tau$) can significantly increase the Fano factor for initial positions close to the target. A similar result holds for the outer sphere. 

Note that one counter-intuitive feature of the resource distribution model is that 
the steady-state mean number of resources in a given target can actually be enhanced by the presence of one or more competing targets. This is illustrated in Fig. \ref{fig7}, where we plot the accumulation rate $\lambda_{1}$ in the inner sphere ($d=3$) as a function of $\tau_{\rm cap}$ and increasing radius $R_2$ of the outer sphere. It can be seen that reducing competition by increasing $R_2$ can decrease $\lambda_{1}$. On the other hand, increasing the time-cost following delivery of resources by increasing $\tau_{\rm cap}$, say, reverses the effects of competition.

\setcounter{equation}{0}
\section{Diffusive search in a wedge domain}

\begin{figure}[b!]
\begin{center} 
\includegraphics[width=6cm]{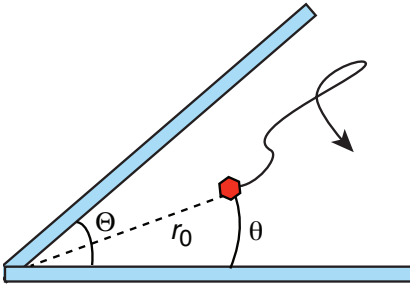} 
\caption{A two-dimensional wedge with opening angle $\Theta$ and absorbing boundaries.}
\label{fig8}
\end{center}
\end{figure}

\begin{figure*}[t!]
\begin{center} 
\includegraphics[width=15cm]{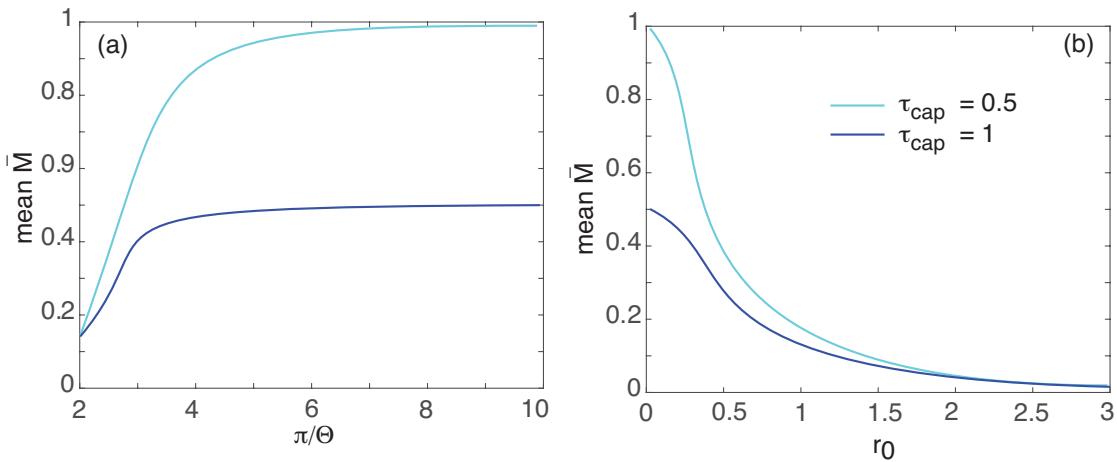} 
\caption{Diffusive search-and-capture in a wedge domain of angle $\Theta$. (a) Plot of the steady-state mean $ \overline{M}$ as a function of $\Theta$ for various $\tau_{\rm cap}$ and $r_0=1$. (b) Corresponding plots as a function of initial radial coordinate $r_0$ for $\Theta=\pi/2$. Other parameter values are $\gamma=1$, $\beta=1$ and $D=1$.}
\label{fig9}
\end{center}
\end{figure*}

\begin{figure*}[t!]
\begin{center} 
\includegraphics[width=15cm]{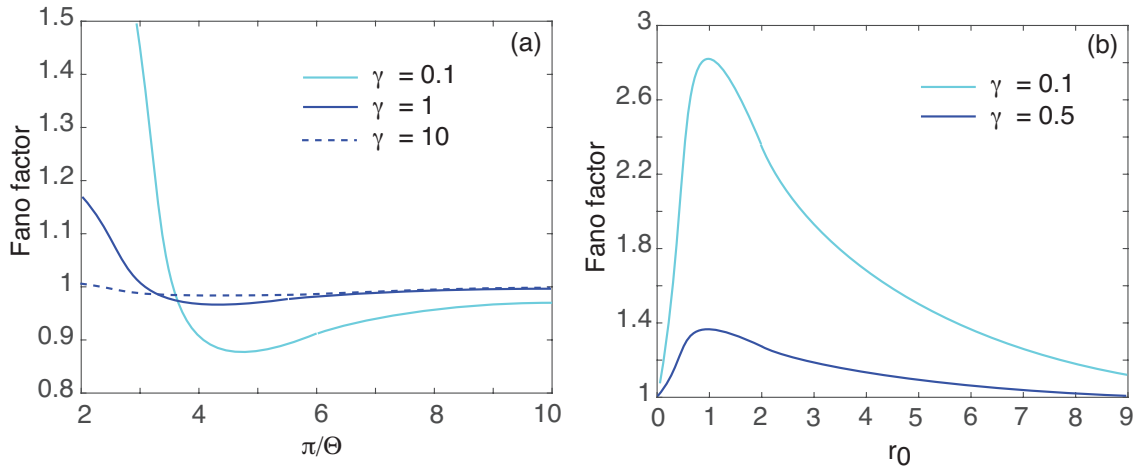} 
\caption{Diffusive search-and-capture in a wedge domain of angle $\Theta$. (a) Plot of the Fano factor as a function of $\Theta$ for various degradation rates $\gamma$ and $r_0=1$. (b) Corresponding plots as a function of initial radial coordinate $r_0$ for $\Theta=\pi/2$. Other parameter values are $\tau_{\rm cap}=1$, $\beta=1$ and $D=1$}
\label{fig10}
\end{center}
\end{figure*}

As our second example, consider a particle diffusing in a two-dimensional wedge domain that subtends an angle $\Theta$, see Fig. \ref{fig8}. The diffusion equation for $p=p(r,\theta,t)$ in polar coordinates takes the form
\begin{equation}
\frac{\partial p}{\partial t}=D\left (\frac{\partial^2p}{\partial r^2}+\frac{1}{r}\frac{\partial p}{\partial r}+\frac{1}{r^2}\frac{\partial^2p}{\partial \theta^2}\right ),
\end{equation}
with $0<r <\infty$ and $0< \theta <\Theta$. The absorbing boundary conditions take the form 
\begin{equation}
p(r,0,t)=0=p(r,\Theta,t),\quad 0<r <\infty.
\end{equation}
Analogous to the previous example, we assume that the initial position of the particle is taken has a fixed radial coordinate $r_0$ and an angle $\theta_0$ this is randomly chosen with a probability density weighted by the factor $\sin (\pi \theta/\Theta)$. That is,
\begin{align}
p(r,\theta,0)=\pi \sin (\pi\theta/\Theta)\frac{\delta(r-r_0)}{2\Theta r_0}.
\end{align}
This means that the particle is equally likely to be captured by either boundary, and the Fourier decomposition of the solution reduces to a single term \cite{Redner}. Laplace transforming the diffusion equation and using separation of variables we find that
\begin{equation}
\widetilde{p}(r,\theta,s)=R(r,s)\sin(\pi \theta/\Theta),
\end{equation}
with $R$ satisfying
\begin{equation}
\frac{d^2R}{dr^2}+\frac{1}{r}\frac{dR}{dr}- \left (\frac{s}{D}+\frac{\nu^2}{r^2}\right )R=-
\nu\frac{\delta(r-r_0)}{2D r_0},
\end{equation}
and $\nu=\pi/\Theta$. It follows that $R$ is a one-dimensional Green's function, which can be calculated using standard methods \cite{Redner}:
\begin{equation}
R(r,s)=\left \{\begin{array}{cc} \frac{\nu}{2D}I_{\nu}(\sqrt{s/D}r)K_{\nu}(\sqrt{s/D}r_0), & r<r_0\\  \\ \frac{\nu}{2D}I_{\nu}(\sqrt{s/D}r_0)K_{\nu}(\sqrt{s/D}r), & r>r_0 \end{array}.
\right .
\end{equation}

The Laplace transform of the probability fluxes into the boundaries at $\theta =0,\Theta$ are equal:
\begin{subequations}
\begin{align}
\widetilde{J}_{0}(r,s)&=\left . \frac{D}{r}\frac{\partial \widetilde{p}(r,\theta,s)}{\partial \theta}\right |_{\theta =0}=\frac{D\nu}{r}R(r,s),\\
\widetilde{J}_{\Theta}(r,s)&=-\left . \frac{D}{r}\frac{\partial \widetilde{p}(r,\theta,s)}{\partial \theta}\right |_{\theta =\Theta}=\frac{D\nu}{r}R(r,s).
\end{align}
\end{subequations}
The total flux into each boundary is thus
\begin{align}
\widetilde{J}(s)&=D\nu \int_0^{\infty} \frac{R(r,s)}{r}dr\\
&=\frac{\nu^2}{2} \bigg( \int_0^{r_0} \frac{I_{\nu}(\sqrt{s/D}r)K_{\nu}(\sqrt{s/D}r_0)}{r}dr
\nonumber \\
&\qquad +\int_{r_0}^{\infty} \frac{I_{\nu}(\sqrt{s/D}r_0)K_{\nu}(\sqrt{s/D}r)}{r}dr\bigg ).\nonumber 
\end{align}
Identifying $\widetilde{J}(s)$ with the Laplace transformed conditional FPT density $\widetilde{f}(s)$ for each boundary, we can use Eqs. (\ref{mean}) and (\ref{FF}) to show that the steady-state mean and Fano factor of resource accumulation in each target boundary are
\begin{equation} \label{wmean}
\overline{M}  = \frac{1}{2\gamma(T+\tau_{\rm cap})},
\end{equation}
and
\begin{equation}
\label{wFF}
 FF= 1+ \frac{\widetilde{\rho}(\gamma)\widetilde{J}(\gamma) }{1-2\widetilde{\rho}(\gamma)\widetilde{J}(\gamma)}-\overline{M}. 
 \end{equation}
 We have used the fact that for two identical targets,
 $s\widetilde{Q}(s)=1-2\widetilde{J}(s)$, see Eq. (\ref{qLT}). In Fig. \ref{fig9} we show plots of the mean $\overline{M}$ in one of the targets as a function of $\nu=\pi/\Theta$ and $r_0$. As expected, increasing the wedge angle $\Theta$ (decreasing $\nu$) reduces the steady-state mean due to the fact that, on average, the particle has further to travel in order to hit a boundary. In the limit $\Theta \rightarrow 0$, $\overline{M}\rightarrow 1/(2\gamma \tau_{\rm cap})$, since $T\rightarrow 0$. Similarly, the mean is a decreasing function of the initial radial coordinate $r_0$. Plots of the Fano factor as a function of $\nu$ and $r_0$ are shown in Fig. \ref{fig10}. In common with the previous example of concentric spheres, the Fano factor exhibits a non-trivial dependence on model parameters, including non-monotonicity and crossover behavior.
Moreover, the qualitative dependence on $\Theta$ differs significantly from the dependence on $r_0$, even though both yield similar behaviors in the case of the mean. This indicates that the non-trivial nature of the higher-order statistics of resource accumulation.

\setcounter{equation}{0}
\section{Single side-branch network}

\begin{figure}[t!]
\begin{center} 
\includegraphics[width=7cm]{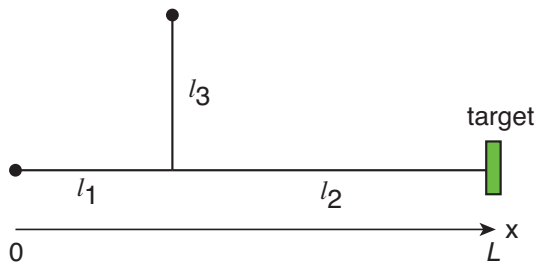} 
\caption{Single side-branch network consisting of two backbone segments of lengths $\ell_1,\ell_2$ and a side branch of length $\ell_3$. There is a single target at one terminal and reflecting boundary conditions at the other two terminals.}
\label{fig11}
\end{center}
\end{figure}

Another important class of FPT problems concerns diffusion on networks. Here we consider the simple configuration considered in \cite{Redner} and shown in Fig. \ref{fig11}. A particle diffuses on two backbone segments $x\in I_1=(0,\ell_1)$ and $x\in I_2 =(\ell_1,\ell_1+\ell_2)$ with a side branch of length $\ell_3$ attached at $x=\ell_1$. Let $p_j$ denote the probability density on the segment of length $\ell_j$, with
\begin{subequations}
\begin{equation}
\label{p1}
\frac{\partial p_j}{\partial t}=D\frac{\partial^2p_j}{\partial x^2}=-\frac{\partial J_j}{\partial x},\quad x\in I_j,\quad j=1,2,
\end{equation}
and
\begin{equation}
\label{p3}
\frac{\partial p_3}{\partial t}=D\frac{\partial^2p_3}{\partial y^2}=-\frac{\partial J_3}{\partial y},\quad y\in (0,\ell_3).
\end{equation}
These are supplemented by the boundary conditions
\begin{equation}
\frac{\partial p_1}{\partial x}(0,t)=0,\quad p_2(L,t)=0,\quad \frac{\partial p_3}{\partial y}(\ell_3,0)=0.
\end{equation} 
where $L=\ell_1+\ell_2$. That is, the terminals at $x=0$ and $y=\ell_3$ are reflecting, while the terminal at $x=L$ is absorbing. Continuity of probability and conservation of probability flux at the junction are ensured by imposing the additional conditions
\begin{align}
p_1(\ell_1,t)&=p_2(\ell_1,t)=p_3(0,t), \\ J_1(\ell_1,t)&=J_2(\ell_1,t) +J_3(0,t).
\end{align}
\end{subequations}
Suppose that a searcher starts at $x=0$ and consider the FPT to reach the target at $x=L$. The FPT density $f(t)$ is identical to the flux $J_2(L,t)$ through $x=L$. The latter can be calculated in Laplace space and one finds that \cite{Redner}
\begin{align}
\label{ftil}
&\widetilde{f}(s)=\bigg (\cosh(\sqrt{s/D}L)\\
&\qquad +\cosh(\sqrt{s/D}\ell_1)\sinh(\sqrt{s/D}\ell_2)\tanh(\sqrt{s/D}\ell_3)
\bigg )^{-1}.\nonumber
\end{align}
Moreover, Taylor expanding $\widetilde{f}(s)$ in $s$ yields the moments of the FPT density. In particular,
\begin{equation}
T=\frac{1}{D}(L^2/2+\ell_2\ell_3).
\end{equation}

\begin{figure}[b!]
\begin{center} 
\includegraphics[width=8cm]{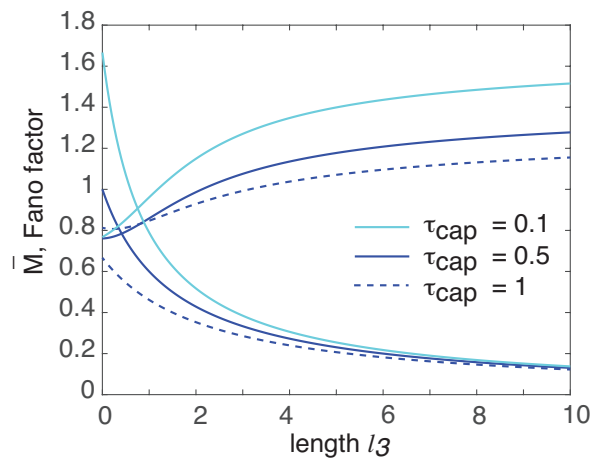} 
\caption{Diffusive search-and-capture on the side branched network, with $\ell_1=1/3$, $\ell_2=2/3$ and variable length $\ell_3$. Plot of the steady-state mean $ \overline{M}$ (monotonically decreasing curves) and Fano factor as a function of the side branch length $\ell_3$ and various delays $\tau_{\rm cap}$. Other parameter values are $\gamma=1$, $\beta=1$ and $D=1$.}
\label{fig12}
\end{center}
\end{figure}

\begin{figure*}[t!]
\begin{center} 
\includegraphics[width=15cm]{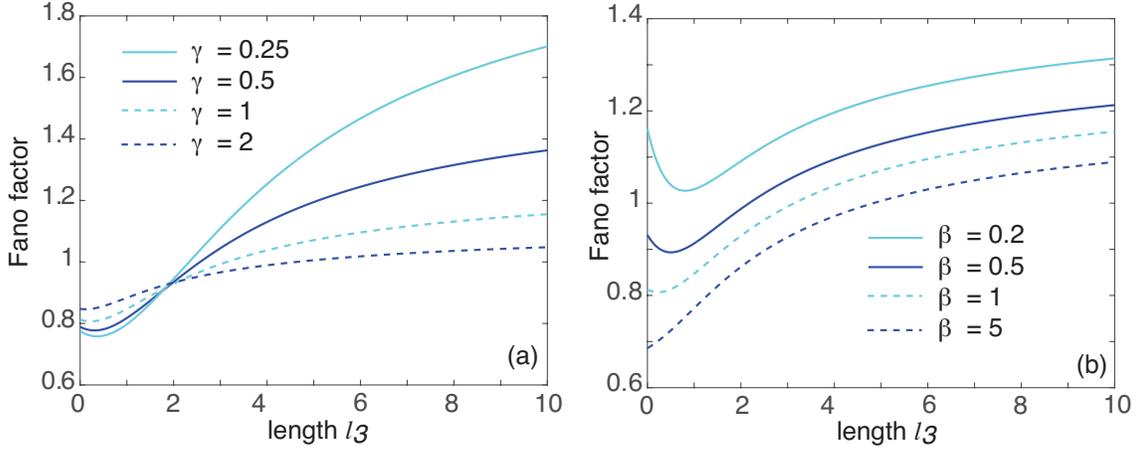} 
\caption{Plot of steady-state Fano factor of the side-branched network as a function of the length $\ell_3$ for (a) various degradation rates $\gamma$ and $\beta=1$; (b) various delay parameters $\beta$ and $\gamma=1$. Here $\tau_{\rm cap}=1$ and other parameters are as in Fig. \ref{fig12}.}
\label{fig13}
\end{center}
\end{figure*}

\begin{figure*}[t!]
\begin{center} 
\includegraphics[width=15cm]{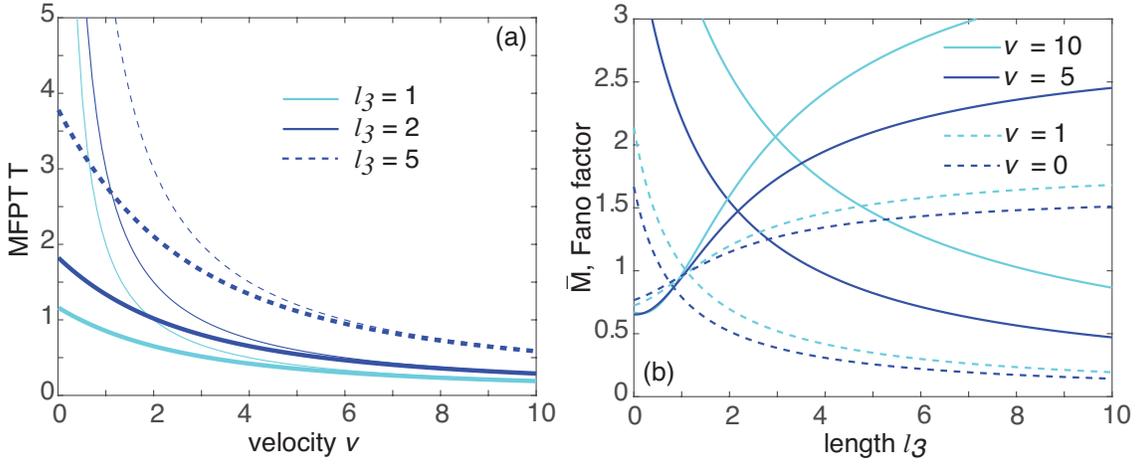} 
\caption{Side branched network with $\ell_1=1/3$, $\ell_2=2/3$ and variable length $\ell_3$. Particle now has an advective component with speed $v$ on the backbone segments. (a) Plot of the MFPT $T$ (thick curves) as a function of $v$ for various lengths $\ell_3$ and $D=1$. The thin curves represent the curves $\sum_{j=1,2,3}\ell_j/v$. (b) Plot of the steady-state mean $ \overline{M}$ (monotonically decreasing curves) and Fano factor as a function of the side branch length $\ell_3$ and various speeds $v$. Other parameter values are $\gamma=1$, $\tau_{\rm cap}=0.1$, $\beta=1$ and $D=1$.}
\label{fig14}
\end{center}
\end{figure*}

Now suppose that there is a build up of resources at the target due to multiple rounds of search and capture. The steady-state mean is simply given by 
\begin{equation}
\label{Mside}
\overline{M} =\frac{1}{\gamma(T+\tau_{\rm cap})},
\end{equation}
and thus vanishes in the limit $\ell_3\rightarrow \infty$ since $T\rightarrow \infty$. Here we focus on the effect of the side branch on the steady-state Fano factor, which takes the form
\begin{equation}
\label{FFside}
 FF= 1+ \frac{\widetilde{\rho}(\gamma)\widetilde{f}(\gamma) }{1-\widetilde{\rho}(\gamma)\widetilde{f}(\gamma)}-\overline{M}. 
 \end{equation}
 We have used the fact that for a single target, $s\widetilde{Q}(s)=1-\widetilde{J}(s)$, see Eq. (\ref{qLT}).
In contrast to the mean, the Fano factor tends to be an increasing function of $\ell_3$, approaching a non-zero constant value in the limit $\ell_3\rightarrow \infty$, as illustrated in Fig. \ref{fig12} for the delay distribution (\ref{gam}). In certain parameter regimes, the Fano factor is a unimodal function of $\ell_3$ with a unique minimum, as illustrated in Fig. \ref{fig13}. We also find that the Fano factor is a decreasing function of $\gamma$ for large $\ell_3$ and an increasing function of $\gamma$ for small $\ell_3$. An analogous crossover phenomenon occurred in the case of concentric spheres, see Fig. \ref{fig6}(a). Finally, as expected, the Fano factor is a decreasing function of $\beta$ for all $\ell_3$, that is, smaller delay fluctuations reduce the Fano factor.

So far we have taken the search process to be unbiased. A directional bias can be including by assuming that when the particle diffuses along the two backbone segments it also has an advective component, whereas pure diffusion occurs along the side branch \cite{Redner}. Eqs. (\ref{p3}) then become
\begin{equation}
\label{pv1}
\frac{\partial p_j}{\partial t}=-v\frac{\partial p_j}{\partial x}+D\frac{\partial^2p_j}{\partial x^2} \quad x\in I_j,\quad j=1,2,
\end{equation}
with $-vp_1(0,t)+Dp_1'(0,t)=0$ and $p_2(L,t)=0$. It is still possible to obtain an analytic expression for the Laplace transformed FPT density, which takes the form (see appendix A)
\begin{widetext}
\begin{align}
\label{J2}
\widetilde{f}(s)=-\frac{(v^2/D+4s)\, \e^{v \ell_1/D}}{su(\ell_1)u(-\ell_2)+\sqrt{sD}\tanh(\sqrt{s/D}\ell_3) u'(\ell_1)u(-\ell_2)+Du'(\ell_1)v(-\ell_2)},\quad \mu_{\pm}=\frac{v\pm \sqrt{v^2+4Ds}}{2D},
\end{align}
\end{widetext}
with
\[u(\ell)=\e^{\mu_+ \ell}-\e^{\mu_-\ell},\quad v(\ell)=\mu_- \e^{\mu_+ \ell}-\mu_+\e^{\mu_-\ell}.
\]
It can be checked that Eq. (\ref{ftil}) is recovered in the limit $v\rightarrow 0$.
The corresponding MFPT is $T=-\widetilde{f}'(0)$, and plots of $T$ as a function of $\ell_3$ are shown in Fig. \ref{fig14}(a). Note that $T\rightarrow \sum_{j=1}^3\ell_j/D$ in the limit $v\rightarrow \infty$.
 As noted in Ref. \cite{Redner}, this asymptotic limit is an example of the so-called equal-time theorem for transport on networks. The latter states that in the limit of large bias, the contribution from each branch of the network is proportional to its length (or volume).

In Fig. \ref{fig14}(b) we show example plots of the mean $\overline{M}$ and Fano factor as a function of $\ell_3$ and various speeds $v$, following multiple rounds of search and capture. As expected, the mean is an increasing function of $v$ and a decreasing function of $\ell_3$. On the other hand, the Fano factor varies with $v$ in an analogous fashion to its variation in $1/\gamma$, see Fig. \ref{fig13}(a). That is, it shows crossover behavior as $\ell_3$ is increased from zero and asymptotes to a constant value in the limit $\ell_3\rightarrow\infty$ that is an increasing function of $v$.

\setcounter{equation}{0}\section{Advection-diffusion on a Cayley tree}

\begin{figure}[b!]
  \centering
  \includegraphics[width=8cm]{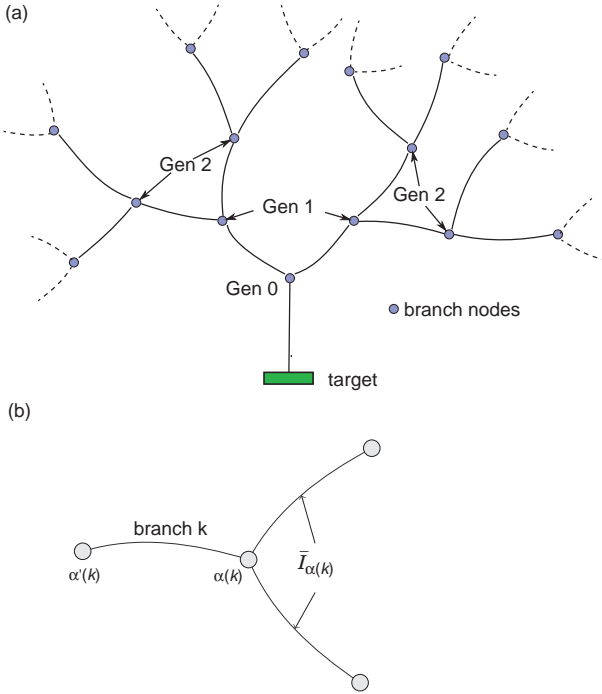}
  \caption{(a) Semi-infinite Cayley tree with coordination number $z=3$ and an absorbing target at the primary node. (b) A branch node
    $\alpha(k)$ is shown in relation to the neighboring branch
    node  $\alpha'(k)$ closest to the primary node.  The
    branch segments extending out from $\alpha(k)$ in the positive
    direction together comprise the set $\neb_{\alpha(k)}$.  }
  \label{fig:notation}
\end{figure}

As our final example, consider advection diffusion on a semi-infinite Cayley tree $\Gamma$ with coordination number $z$. The example of $z=3$ is shown in Fig. \ref{fig:notation}(a).
We will assume for simplicity that the drift velocity $v$, diffusivity $D$, and
branch length $L$ are identical throughout the tree so that we can exploit the recursive nature of the infinite tree. Suppose that there is a target at the terminal (primary) node of the tree. We will calculate the Laplace transform of the flux through the target using the iterative method introduced in Ref. \cite{Newby09}. This will then allow us to explore how the steady-state mean and variance of the number of resources at the target depend on parameters of the model.

Denote the first branch node opposite the terminal node
by $\alpha_{0}$. For every other branching node $\alpha \in
\Gamma$ there exists a unique direct path from $\alpha_0$ to
$\alpha$ (one that does not traverse any line segment more than
once). We can label each node $\alpha \neq \alpha_0$ uniquely by the
index $k$ of the final segment of the direct path from $\alpha_0$ to
$\alpha$ so that the branch node corresponding to a given 
segment label $k$ can be written $\alpha(k)$. We denote the other node of
segment $k$ by $\alpha'(k)$. Taking the primary branch to be $k=0$, it follows that $\alpha(0)=\alpha_0$ and $\alpha'(0)$ is the terminal node. We also
introduce a direction on each segment of the tree such that every
direct path from $\alpha'(0)$ always moves in the positive
direction. Consider
a single branching node $\alpha \in \Gamma $ and label the set of
segments radiating from it by $\branch_{\alpha}$. Let $\neb_{\alpha}$ denote the set of $z-1$ line segments
$k\in\branch_{\alpha}$ that radiate from $\alpha $ in a
positive direction, see Fig. \ref{fig:notation}(b).
Using these various
definitions we can introduce the idea of a generation. Take $\alpha_0$
to be the zeroth generation. The first generation then consists of the
set of nodes $\Sigma_1=\{\alpha(k),k\in \neb_{\alpha_0}\}$, the second
generation is $\Sigma_2=\{\alpha(l),l\in \neb_{\alpha},\alpha\in
\Sigma_1\}$ etc.

Denote the position coordinate
along the $i$-th line segment by $x$, $0 \leq x \leq L$, where $L$
is the length of the segment and $0\leq i< \infty$. Given the above labeling scheme for nodes, we take $x(\alpha'(i))=0$ and $x(\alpha(i))=L$. Let $p_i$ denote
the probability density on the $i$-th segment, which evolves according
to the Fokker-Planck equation
\begin{eqnarray}
\label{FPb}
  \frac{\partial p_i}{\partial t} = D\frac{\partial^2 p_i}{\partial
    x^2}-v\frac{\partial p_i}{\partial x} ,
  \quad 0 < x < L.
\end{eqnarray} 
As a further simplification, we assume that the searcher initiates its search on the primary branch so that
\begin{equation}
p_i(x,0|x_0)= \delta(x-x_0)\delta_{i,0},\quad 0 < x_0<L.
\end{equation}
This initial condition means that all branches of a given generation are equivalent.
Let $\flx[p_i]$ denote the corresponding probability current or flux, which is taken to be positive in the direction flowing away from the primary node at the soma:
\begin{equation}
  \flx[p]\equiv  - D\frac{\partial p}
        {\partial x} +v p.
\end{equation}
The open boundary condition on the primary branch is $p_0(0,t|x_0) =0$.
At all branch nodes $\alpha\in \Sigma_n$ of the $n$-th generation we impose the
continuity conditions
\begin{equation}
\label{cont}
  p_i(x(\alpha),t|x_0) = \Phi_{n}(x_0,t) ,\quad \mbox{for all}\ i \in {\mathcal I}_{\alpha},\quad \alpha\in \Sigma_n,
\end{equation}
where the $\Phi_{n}(x_0,t)$ are unknown functions, which will ultimately be determined by imposing current conservation at each branch node:
\begin{equation}
\label{cons}
   \sum_{i \in{\mathcal I}_{\alpha}} \flx[p_i(x(\alpha),t|x_0)] = 0.
\end{equation}
Note that for the upstream segment $j\notin \overline{\mathcal I}_{\alpha}$, $x(\alpha)=L$ and the corresponding flux $\flx[p_j(L,t|x_0)]$ flows into the branch node, whereas for the remaining $z-1$ downstream segments $k\in 
\overline{\mathcal I}_{\alpha}$ we have $x(\alpha)=0$ and the flux $\flx[p_k(0,t|x_0)]$ flows out of the branch node.

\subsection{Calculation of flux at terminal node}

After Laplace transforming the above system of equations on the Cayley tree, we obtain the following system of equations for any $i$ such that $\alpha(i)\in \Sigma_n$:
\begin{equation}
\label{AV}
\left [D\frac{\partial^{2} }{\partial x^{2}} 
     -v\frac{\partial}{\partial x} -s \right ]\tu_{i}(x,s) = - \delta_{i,0}\delta(x-x_0).
\end{equation}
together with the boundary conditions
\begin{equation}
\tu_i(0,t) = \widetilde{\Phi}_{n-1}(s),\quad  \tu_i(L,t) = \widetilde{\Phi}_{n}(s).
\end{equation}
Note that $  \widetilde{\Phi}_{-1}= 0$. (For notational concvenience, we drop the explicit dependence on the initial position $x_0$.)
The solution in each branch is given by the corresponding
finite interval Green's function ${\mathcal G}$ with homogeneous boundary conditions supplemented by terms satisfying the boundary conditions. That is,
  \begin{align}
  \label{GB}
   \tu_i(x,s) &=-  \delta_{i,0}{\mathcal G}(x,x_0;s) +
    \widetilde{\Phi}_{n-1}(s)\widehat{F}(x,s) \nonumber \\
    &\quad + \widetilde{\Phi}_{n}(s)F(x,s)
  \end{align}
for $\alpha(i)\in \Sigma_n$.
The Green's function ${\mathcal G}$ satisfies
\begin{equation}
\label{AV2}
\left [D\frac{\partial^{2} }{\partial x^{2}} 
     -v\frac{\partial}{\partial x} -s \right ]{\mathcal G}(x,y;s) = \delta(x-y),
\end{equation}
with ${\mathcal G}(0,y;s)=0={\mathcal G}(L,y;s)$.
Using standard methods, the Green's function is given by
\begin{equation}
  \label{eq:finite_gf_oo}
 {\mathcal G}(x,y;s) = \left \{
  \begin{array}{cc}
    \frac{\displaystyle \uoo(x,s)\uoo(y-L,s)}{\displaystyle DW(s) },&0\leq x \leq y \\ \\
    \frac{\displaystyle \uoo(x-L,s)\uoo(y,s)}{\displaystyle  DW(s) }, &y\leq x \leq L
  \end{array} \right . ,
\end{equation}
where
\begin{align}
\label{bag}
  \uoo(x,s) &= \e^{\mu_+(s)x} - \e^{\mu_-(s)x},\quad \mu_{\pm}=\frac{v\pm \sqrt{v^2+4Ds}}{2D},
  \end{align}
  and $W$ is the Wronskian
  \begin{align}
  W(s)  &= \uoo(y,s)\uoo'(y-L,s)-\uoo'(y,s)\uoo(y-L,s),
\end{align}
which is independent of $y$.
The functions
$F(x,s)$ and $\widehat{F}(x,s)$ satisfy the homogeneous version of Eq. (\ref{AV}) with boundary conditions $F(0,s)=0,F(L,s)=1$ and $ \widehat{F}(0,s)=1,\widehat{F}(L,s)=0$:
\begin{align}
\label{Fk}
  F(x,s)=\frac{\psi(x,s)}{\psi(L,s)},\quad   \widehat{F}(x,s)=\frac{\psi(x-L,s)}{\psi(-L,s)}.
\end{align}

The unknown functions $\Phi_{n}$ are determined by imposing the
current conservation condition \eqref{cons} at each branch node and using the identity
$\flx[\Phi_{n} F]=\Phi_{n} \flx[F]$, which follows from the observation that $\Phi_{n}$ is $x$--independent. At the zeroth generation node
${\alpha_{0}}$, the current conservation equation is given by (suppressing the $s$ variable)
  \begin{align}
    & \Phi_{0} \flx[F](L)\\
    &\quad = (z-1)
    \Phi_{1}\flx[F](0) +(z-1)\Phi_0\flx [\widehat{F}](0) + \targT_{0},  \nonumber  
  \end{align}
  where 
\begin{equation}
 \targT_{0}\equiv  \flx[{\mathcal G}](L),
\end{equation}
and at all  branching nodes $\alpha \in\Sigma_{n}$, $1\leq n$ we
have
\begin{align}
    &\Phi_{n-1}\flx[\widehat{F}](L)
    +\Phi_{n}\flx[F](L) 
    =(z-1)\Phi_{n+1}\flx[F](0)\nonumber\\
    &\qquad +(z-1) \Phi_{n}
    \flx[\widehat{F}](0).
  \end{align}
Note that $\targT_{0}$ depends on the source location $x_0$ through its dependence on the finite interval Green's function ${\mathcal G}$; this then generates an $x_0$--dependence of the functions $\Phi_{\alpha}$

The four possible contributions to the probability flux at any branch $k\neq 0$ are
\begin{subequations}
\begin{align}
 g(s)&\equiv \flx[\widehat{F}](L) =\frac{\psi'(0,s)}{\psi(-L,s)}=
  \frac{D\eta(s)\e^{{vL}/{2D}}}
  {\sinh(\eta(s) L)},\\
  h(s)&\equiv\flx[F](L) =\frac{\psi'(L,s)}{\psi(L,s)}=
  -D\eta(s)\coth(\eta(s) L)
  +\frac{v}{2}\\
 \bar{g}(s)&\equiv \flx[F](0) =\frac{\psi'(0,s)}{\psi(L,s)}=
 - \frac{D\eta(s) \e^{{-vL}/{2D}}}
  {\sinh(\eta(s) L)}\\
  \bar{h}(s)&\equiv\flx[\widehat{F}](0)=\frac{\psi'(-L,s)}{\psi(-L,s)}=
  D\eta(s)\coth(\eta(s) L) +\frac{v}{2},
\end{align}
\end{subequations}
where
\begin{equation}
\eta(s)=\frac{\sqrt{v^2+4Ds}}{2D}.
\end{equation}
Using these definitions the current conservation equations simplify to
\begin{equation}
\label{pip}
-H\Phi_{0}
  +G\Phi_{1} =  \targT_{0}
\end{equation}
for the first branch, and
\begin{equation}
\label{itp}
  g\Phi_{n-1}
  -H\Phi_{n}
 +G\Phi_{n+1}
  =   0
\end{equation}
for $n>0$,
where
\begin{equation}
\label{H}
H= (z-1)\bar{h}-h,\quad G=-\bar{g}(z-1).
\end{equation}

\begin{figure}[t!]
\begin{center} 
\includegraphics[width=8cm]{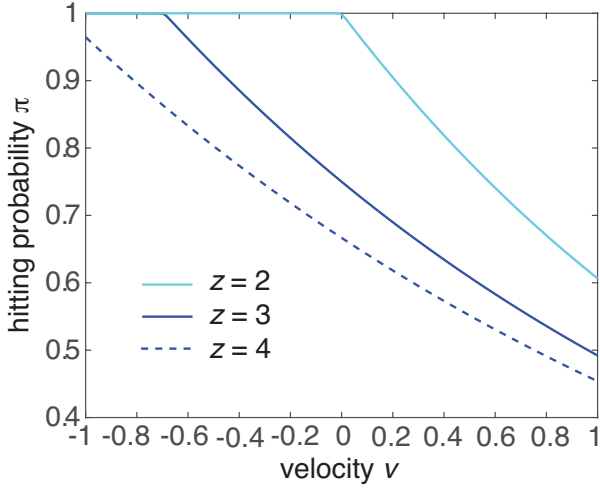} 
\caption{Advection-diffusion on a semi-infinite Cayley tree. Plots of hitting probability $\pi$ as a function of velocity $v$ for different coordination numbers $z$. Other parameters are $D=1$ and $L=1$.}
\label{fig16}
\end{center}
\end{figure}

\begin{figure*}[t!]
\begin{center} 
\includegraphics[width=15cm]{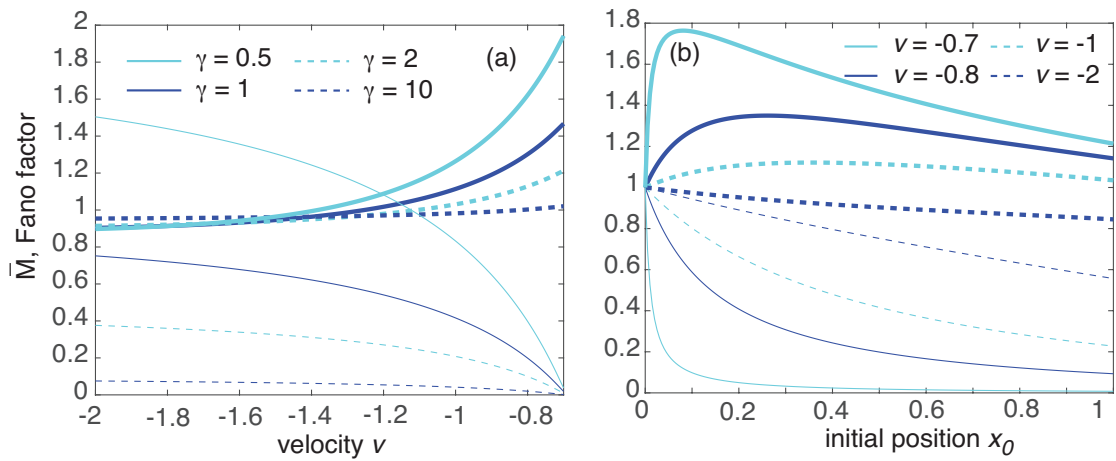} 
\caption{Advection-diffusion on a semi-infinite Cayley tree. (a) Plots of steady state mean (thin curves) and Fano factor (thick curves) as a function of the velocity $v$, $v<-v_c$, for various degradation rates $\gamma$ and $x_0=0.5$. (b) Corresponding plots as a function of the initial position $x_0$ for $\gamma=1$ and various velocities. Other parameters are $D=1$, $L=1$, $z=3$, $\tau_{\rm cap}=1$ and $\beta=1$.}
\label{fig17}
\end{center}
\end{figure*}

 The second-order difference equation can be solved using the ansatz $\Phi_n=\lambda^n\Phi_0$, which yields a quadratic equation for $\lambda$:
 \begin{equation}
 G\lambda^2-H\lambda+g=0.
 \end{equation}
 Taking the smaller root, since the larger root is greater than one and thus does not yield a bounded solution, gives
 \begin{equation}
 \lambda=\frac{1}{2G}\left [H-\sqrt{H^2-4gG}\right ].
 \end{equation}
 It can be checked that $\lambda$ is real since
 \begin{align*}
 H^2-4gG&>\frac{z^2(D\eta)^2\cosh^2(\eta L)}{\sinh^2(\eta L)}-\frac{4(z-1)(D\eta)^2}{\sinh^2(\eta L)}\\
 &=\frac{(D\eta)^2}{\sinh^2(\eta L)}\left (z^2\cosh^2(\eta L)-4(z-1)\right )\\
 &>\frac{(D\eta)^2}{\sinh^2(\eta L)}(z-2)^2>0.
 \end{align*}
In addition, $\lambda <1$ since this inequality is equivalent to the condition
 \begin{align*}
 H-\sqrt{H^2-4gG}<2G,
 \end{align*}
 which can be rewritten as
 \[(H-2G)^2 <H^2-4gG \implies g+G<H,\]
 and $g+G<H$.
  We now substitute the solution for $\Phi_1$ into Eq. (\ref{pip}) and rearrange to obtain the following expression for $\widetilde{\Phi}_0$:
 \begin{equation}
\widetilde{\Phi}_0=\frac{\chi_0}{H-\lambda G}.
\end{equation}
Having determined $\widetilde{\Phi}_0$, the Laplace transform of the flux into the terminal node takes the explicit form (after reincorporating the dependence on the initial position)
\begin{align}
\label{Jtree}
\widetilde{J}(x_0,s)&=D\frac{\partial \widetilde{p}_0}{\partial x}(0,s|x_0)=-\left . D\frac{\partial {\mathcal G}(x,x_0;s)}{\partial x} \right |_{x=0}\nonumber \\
&\quad +D\left . \widetilde{\Phi}_{0}(x_0,s)\frac{\partial F(x,s)}{\partial x}\right |_{x=0}.
\end{align}

The function $\widetilde{J}(x_0,s)$ determines the hitting probability $\pi$ and corresponding conditional MFPT $T$ according to
\begin{equation}
\pi(x_0) = \widetilde{J}(x_0,0),\quad \pi(x_0) T(x_0)=-\widetilde{J}'(x_0,0).
\end{equation}
Note that unbiased one-dimensional diffusion is transient in one dimension ($z=2$) and on a Cayley tree. This means that $\pi <1$ for $v\geq 0$. On the other hand, there exists a critical negative (inward) velocity $v_c$, $v_c<0$, such that $\pi =1$ for $v<v_c$. In other words, there is a critical phase transition from recurrent to transient transport. This transition point is identical to the so-called localization-delocalization threshold \cite{Bressloff97}. In order to define the latter, suppose that the particle starts at the terminal node, $p_i(x,0)=\delta_{i,0}\delta(x)$. The initially localized probability density will tend to diffuse away from that origin, but this is counteracted by an inward velocity field on the tree. If, in steady state, the concentration remaining
at the origin has not decayed to zero, we say the
system is localized, otherwise it is delocalized. By studying the steady-state solution it can be shown that $v_c=-\ln(z-1)$. Hence $v_c=0$ for $z=2$ (a semi-infinite line), $v_c\approx -0.69$ for $z=3$ and $v_c\approx -1.1$ for $z=4$.
The existence of the recurrent-transient transition for the hitting probability is confirmed in in Fig. \ref{fig16}. Note that for $v>v_c$, the hitting probability is a decreasing function of both $v$ and $z$.

\subsection{Mean and Fano factor of resource distribution} 

Suppose that $z=3$ and $v<v_c$ so that the particle eventually finds the target with unit probability. We can then substitute Eq. (\ref{Jtree}) for $\widetilde{J}(s)$ into Eqs. (\ref{Mside}) and (\ref{FFside}) in order to determine the steady-state mean and Fano factor for the number of resources delivered to the target. In Fig. \ref{fig17}(a) we plot the mean and Fano factor as a function of the velocity and various degradation rates for $z=3$. (Qualitatively similar results occur for other coordination numbers $z$.) The mean is a monotonically decreasing function of $v$ and $\overline{M}\rightarrow 0$ as $v\rightarrow v_c^-$ (due to $T\rightarrow \infty$). On the other hand, the Fano is an increasing function of $v$. In Fig. \ref{fig17}(b) we plot the mean and Fano factor as a function of the initial position for various velocities. As expected $\overline{M}$ decreases with increasing distance $x_0$ from the target. In this case, the Fano factor is a non-monotonic function of $x_0$ for a range of velocities, with the maximum increasing significantly as $v\rightarrow v_c^-$.

\section{Discussion}

In this paper we investigated target resource accumulation under multiple rounds of diffusive search-and-capture events. Each event involved a FPT problem to reach one or more absorbing boundaries (targets) of some specified search domain. The steady-state mean and Fano factor of the number of resources were expressed in terms of the Laplace transformed probability fluxes into the target(s). We considered a number of classical search problems that involved relatively simple geometries: concentric spheres, a wedge domain, a side-branched network (simple comb), and a semi-infinite Cayley tree. This allowed us to explicitly calculate the probability fluxes in Laplace space. A complementary approach is to consider a more general search domain ${\mathcal U}$ with one or more interior boundaries $\partial \calU_k$ along the lines of Fig. \ref{fig1}. One can then used asymptotic analysis to investigate the distribution of resources in the small target limit \cite{Bressloff20A}.

One motivation for the target-centric perspective taken in this paper is the intracellular transport of proteins and lipids to the cell membrane and subcellular compartments such as the cell nucleus, endoplasmic reticulum, and synapses along the axons of neurons \cite{Bressloff20a}. For example, within the context of Fig. \ref{fig1}, $\calU$ could be identified with the cell cytoplasm, $\partial \calU_2$ with the cell membrane, and $\partial \calU_1$ with the cell nucleus, say. Similarly, advection-diffusion on a tree-like structure (section 6) has applications to the motor-driven transport of vesicles in the dendrites of neurons. Although active transport is typically modeled in terms of velocity jump processes, it is possible to reduce the latter to an effective advection-diffusion process \cite{Newby09}. However, in order to further develop specific applications, it would be necessary to consider in more detail the active mechanism by which a new search-and-capture process is started following target capture. (An analogous issue concerns single search-and-capture processes with stochastic resetting \cite{Evans20}.) In the examples, we assumed that the main contribution to the delays between events arose from the unloading/loading of resources, rather than the time for a searcher to return to its initial loading position. In addition, for the first two examples, we chose a distribution of initial positions in order to exploit symmetries of the underling geometry; radial symmetry in the case of concentric spheres. Elsewhere, we consider more realistic resetting-after-capture processes in the case of one-dimensional search processes \cite{Bressloff20a}.

Another issue that is not addressed in this paper is how resources are distributed within a spatially extended target such as the spherical boundaries in section III. If the latter are interpreted as cellular membranes, then lateral diffusion within each membrane would lead to a uniform distribution of resources. However, it is also possible that resources are localized to specific subdomains of the membrane, which would mean partitioning the boundary into multiple target domains.

The particular example of advection-diffusion on a Cayley tree (section VI) also has a number of possible generalizations. First, in the analysis of localization-delocalization transitions, more general results for the phase transition were obtained by considering Cayley trees with quenched disorder in the distribution of branch velocities \cite{Bressloff97}. Second, it is possible to extend the iterative method for solving the Laplace transformed advection-diffusion equation to the case of finite trees \cite{Newby09}. In this case, there exists multiple terminal nodes, each of which could act as a target.

Finally, note that in this paper we focused on a single searcher, whereas a more common scenario is to have many parallel searchers. However, our results carry over to this case provided that the searchers are independent. That is, suppose that there are ${\mathcal N}$ independent, identical searchers. Statistical independence implies that both the steady-state mean and variance scale of resources within a target scale as ${\mathcal N}$. Hence, the Fano factor is independent of ${\mathcal N}$, whereas the coefficient of variation scales as $CV_k\sim 1/\sqrt{\mathcal N}$. The latter indicates that the size of fluctuations decreases as the number of searchers increases, which is also a manifestation of the law-of-large numbers.

\vspace{-0.5cm}

\setcounter{equation}{0}
\renewcommand{\theequation}{A.\arabic{equation}}
\renewcommand{\thesubsection}{A.\arabic{subsection}}
\section*{Appendix A: Calculation of FPT density for side-branched network.}

Write the solution for the Laplace transformed probability density in the three segments as
\begin{align*}
\widetilde{p}_1(x,s)&=A_1\e^{\eta_+x}+B_1\e^{\eta_-x},\quad x\in (0,\ell_1),\\
\widetilde{p}_2(x,s)&=A_2\left (\e^{\eta_+(x-L)}-e^{\eta_-(x-L)}\right )\\
&=A_2u(x-L),\quad x\in (\ell_1,L),\\
\widetilde{p}_3(x,s)&=A_3\cosh(\sqrt{s/D}(\ell_3-y),\quad 0<y<\ell_3.
\end{align*}
We have already imposed the absorbing boundary condition at $x=L$ and the reflecting boundary condition at $y=\ell_3$. The reflecting boundary condition at $x=0$ implies that
\begin{align*}
\frac{\partial \widetilde{p}_1}{\partial x}(0,s)-\frac{v}{D}\widetilde{p}_1(0,s)=-\frac{1}{D},
\end{align*}
that is
\[(\eta_+-v/D)A_1+(\eta_--v/D)B_1=-1/D.\]
Using the identities $\eta_{\pm}-v/D=-\eta_{\mp}$, we find
\begin{equation}
\label{A1}
\eta_-A_1+\eta_+B_1=1/D.
\end{equation}
Continuity of the solution at the junction yields the pair of equations
\begin{equation}
\label{A2}
A_2u(-\ell_2)=A_3\cosh (\sqrt{s/D}\ell_3),
\end{equation}
and
\[A_1\e^{\eta_+\ell_1}+B_1 \e^{\eta_-\ell_1}=A_2u(-\ell_2).\]
Using Eq. (\ref{A1}) to eliminate $B_1$ then gives
\begin{equation}
A_1u'(\ell_1)+D^{-1}\e^{\eta_-L}=\eta_+u(-\ell_2)A_2.
\label{A3}
\end{equation}
Finally, imposing conservation of current at the junction,
\begin{align*}
\frac{\partial \widetilde{p}_1}{\partial x}(\ell_1,s)-\frac{v}{D}\widetilde{p}_1(\ell_1,s)&=\frac{\partial \widetilde{p}_2}{\partial x}(\ell_1,s)-\frac{v}{D}\widetilde{p}_2(\ell_1,s)\\
&\quad +\frac{\partial \widetilde{p}_3}{\partial y}(0,s).
\end{align*} 
Substituting the various solutions gives
\begin{align*}
&(\eta_+-v/D)A_1\e^{\eta_+\ell_1}-(\eta_--v/D)B_1\e^{\eta_-\ell_1}\\
&=
A_2\left [(\eta_+-v/D) \e^{-\eta_+\ell_2}-(\eta_--v/D)\e^{-\eta_-\ell_2}\right ]\\
&\quad -\sqrt{\frac{s}{D}} A_3\sinh(\sqrt{s/D}\ell_3).
\end{align*}
Substituting for $B_1$ and $A_3$ using eqs. (\ref{A1}) and (\ref{A2}), we have
\begin{align}
\label{A4}
&\eta_-u(\ell_1)A_1+D^{-1}\e^{\eta_-\ell_1}\\
&=\left [\sqrt{\frac{s}{D}} \tanh(\sqrt{s/D}\ell_3)u(-\ell_2)+v(-\ell_2)\right ]A_2,\nonumber
\end{align}
with $ v(-\ell_2)=\eta_- \e^{-\eta_+ \ell_2}-\eta_+\e^{-\eta_-\ell_2}$.

We can now use Eqs. (\ref{A3}) and (\ref{A4}) to eliminate $A_1$ and obtain the following expression for $A_2$:
\begin{widetext}
\begin{align}
A_2&=\frac{\e^{\eta_-\ell_1}}{D}\frac{u'(\ell_1)-\eta_-u(\ell_1)}
{\left [\sqrt{s/D}\tanh(\sqrt{s/D}\ell_3)u(-\ell_2)+v(-\ell_2)\right ]u'(\ell_1)-\eta_+\eta_-u(\ell_1)u(-\ell_2)}\nonumber \\
&=\frac{(\eta_+-\eta_-)\e^{[\eta_-+\eta_-]\ell_1}}
{\left [\sqrt{sD}\tanh(\sqrt{s/D}\ell_3)u(-\ell_2)+Dv(-\ell_2)\right ]u'(\ell_1)+su(\ell_1)u(-\ell_2)}.
\end{align}
We have used $\eta_+\eta_-=-s/D$. Finally, noting that 
\begin{align*}
\widetilde{J}_2(L,s)=-D\frac{\partial \widetilde{p}_2}{\partial x}(L,s)=-D(\eta_+-\eta_-) A_2,
\end{align*}
$\eta_++\eta_-=v/D$, and  $\eta_+-\eta_-=\sqrt{v^2+4Ds}/D$, we obtain Eq. (\ref{J2}).

\end{widetext}

\end{document}